# Microarrays denoising via smoothing of coefficients in wavelet domain

M. MASTRIANI, A. E. GIRALDEZ
SAOCOM Mission
National Commission of Space Activities (CONAE)
751 P. Colon Ave., C1063ACH Buenos Aires
ARGENTINA
{mmastri,agiral}@conae.gov.ar   http://www.conae.gov.ar

*Abstract:* - We describe a novel method for removing noise (in wavelet domain) of unknown variance from microarrays. The method is based on a smoothing of the coefficients of the highest subbands. Specifically, we decompose the noisy microarray into wavelet subbands, apply smoothing within each highest subband, and reconstruct a microarray from the modified wavelet coefficients. This process is applied a single time, and exclusively to the first level of decomposition, i.e., in most of the cases, it is not necessary a multirresoltuion analysis. Denoising results compare favorably to the most of methods in use at the moment.

## 1 Introduction

A microarray is affected by noise in its acquisition and processing. Microarray denoising is used to remove the additive noise while retaining as much as possible the important image features. In the recent years there has been an important amount of research on wavelet thresholding and threshold selection for bioimages denoising, e.g., microarray images [1], [2], because wavelet provides an appropriate basis for separating noisy signal from the image signal. The motivation is that as the wavelet transform is good at energy compaction, the small coefficients are more likely due to noise and large coefficient due to important signal features [3]-[5]. These small coefficients can be thresholded without affecting the significant features of the image.

In general, the results of the microarray processing combine two sample images that after further image processing, gene expression data can be produced for further analysis, such as gene clustering or identification [1], [2]. These three crucial steps, experiment, image processing and data analysis, determine the success or not of the microarray analysis. Image processing plays a potentially large impact on the subsequent analysis. In recent years, a large number of commercial tools have been developed in microarray image processing [1], [2]. The tasks of all these tools mainly focus on two major targets, namely: spot segmentation and spot intensity extraction. However, the quality of the images from the experiments is not always perfect. The gene array experiments involve a large number of error-prone steps which lead to a high level of noise in the resulting images [1], [2]. Hence, the accuracy of the gene expressions derived from these images will largely be affected in the process.

In order to assure the accuracy of the gene expression, normally the replicated experiments and incorporated statistical methods are needed to estimate the errors [1], [2]. These methods deal mainly with measurement error, such as preparation of the sample, cross hybridization, and fluctuation of fluorescence value from gene to gene. But none deals particularly with the effect of the noise [1], [2].

In fact, the thresholding technique is the last approach based on wavelet theory to provide an enhanced approach for eliminating such noise source and ensure better gene expression. Thresholding is a simple non-linear technique, which operates on one wavelet coefficient at a time. In its basic form, each coefficient is thresholded by comparing against threshold, if the coefficient is smaller than threshold, set to zero; otherwise it is kept or modified. Replacing the small noisy coefficients by zero and inverse wavelet transform on the result may lead to reconstruction with the essential signal characteristics and with less noise. Since the work of Donoho & Johnstone [5], there has been much research on finding thresholds, however few are specifically designed for images [3], [4], [6].

Unfortunately, this technique has the following disadvantages:

1. it depends on the correct election of the type of thresholding, e.g., OracleShrink, VisuShrink (soft-thresholding, hard-thresholding, and semi-soft-thresholding), SureShrink, Bayesian soft thresholding, Bayesian MMSE estimation, Thresholding Neural Network (TNN), due to Zhang, NormalShrink, , etc. [3]-[7],
2. it depends on the correct estimation of the threshold which is arguably the most important design parameter,
3. it doesn't have a fine adjustment of the threshold after their calculation,
4. it should be applied at each level of decomposition, needing several levels, and
5. the specific distributions of the signal and noise may not be well matched at different scales.

Therefore, a new method without these constraints will represent an upgrade.

## 2 Smoothing of Coefficients (SC) in Wavelet Domain

We decompose the noisy microarray into four wavelet subbands: Coefficients of Approximation (CA), and noisy coefficients of Diagonal Detail ($CDD_n$), Vertical Detail ($CVD_n$), and Horizontal Detail ($CHD_n$), respectively. We apply a bidimensional smoothing within each highest subband, and reconstruct a microarray from the modified wavelet coefficients, that is to say, denoised coefficients of Diagonal Detail ($CDD_d$), Vertical Detail ($CVD_d$), and Horizontal Detail ($CHD_d$), respectively, as shown in Fig. 1, where: DWT-2D is the Bidimensional Discrete Wavelet Transform, and IDWT-2D is the inverse of DWT-2D.

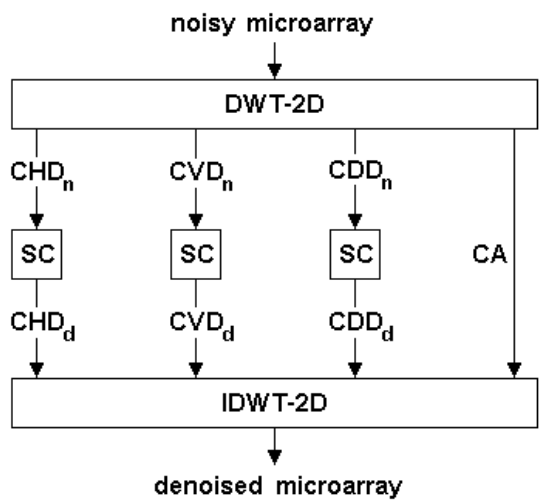

Fig. 1. Smoothing of Coefficients (SC) in wavelet domain.

If we use an original microarray of *R-by-C* pixels, then each subbands will have *(R/2)-by-(C/2)* coefficients. The SC process is applied - in principle - a single time, and exclusively to the first level of decomposition.

On the other hand, to protect the edges from blurring while smoothing the respective coefficients of subband, an appropriate filter must be applied. The most of statistical filters have a speckle reduction approach that performs spatial filtering in a square-moving window defined as kernel, and is based on the statistical relationship between the central pixel and its surrounding pixels as shown in Fig. 2.

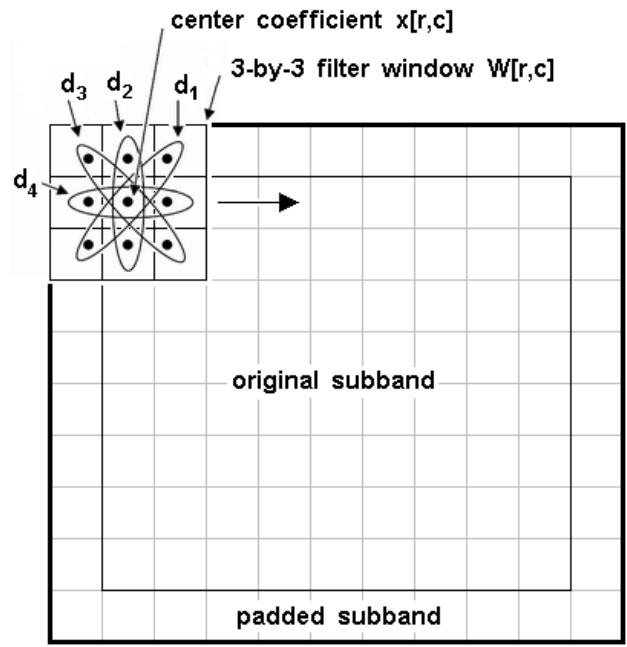

Fig. 2. 3-by-3 filter window for noise smoothing over each highest subband (CHD, CVD, and CDD) using directional smoothing.

The size of the filter window can range from 3-by-3 to 33-by-33, with an odd number of cells in both directions. A larger filter window means that a larger area of the subband will be used for calculation and requires more computation time depending on the complexity of the filter's algorithm. If the size of filter window is too large, the important details will be lost due to over smoothing. On the other hand, if the size of the filter window is too small, speckle reduction may not be very effective. In practice, a 3-by-3 or a 7-by-7 filter window usually yields good results in the cases under study [8]-[9].

For example, if the statistical filter used inside SC method (that is to say, over each highest subbands) is Directional Smoothing (DS), then, let

$x[r, c]$ denote the value of the corresponding noisy detail coefficient at location $(r, c)$. Let $W[r, c]$ represent the group of coefficients contained in a filtering window with the size of 3-by-3 pixels and centered at location $(r, c)$ of the corresponding noisy detail coefficient, as shown in Fig. 2:

$$W[r, c] = \{ x[r+p, c+q] \mid (p, q) \in [-1,0,1] \} \quad (1)$$

where $p$ and $q$ are integer indices each individually ranging from $-1$ to $1$.

Here, $r$ and $c$ are the row and the column indices, respectively, with $2 \leq r \leq (R/2)-1$ and $2 \leq c \leq (C/2)-1$.

The noisy input subband is processed by sliding a 3-by-3 filtering window on the subband. The window is started from the upper-left corner of the subband and moved sideways and progressively downwards in a raster scanning fashion. Meanwhile, the directional averaging filter (selective with respect to direction) examine the average based on several directionally oriented masks, as shown in Fig. 2, and it compute averages in $d_1$, $d_2$, $d_3$ and $d_4$ directions $Avg\ i$; $i = 1, 2, 3, 4$. Such as, $y[r,c] = Avg\ i$ where $Avg\ i$ is the one closest in amplitude to $x(r,c)$, i.e. $|x(r,c)-Avg\ i|$ minimum. That is to say, it has a tendency not to destroy boundaries. On the other hand, the directional analysis can also be used to check if a coefficient belongs to a directional edge (leave unchanged) or is noise (remove noise).

Finally, any used filter performs the filtering based on either local statistical data given in the filter window to determine the noise variance within the filter window, or estimating the local noise variance of the subband under study, e.g., Median, Lee, Kuan, Gamma-Map, Enhanced Lee, Frost, Enhanced Frost [3], [8], Wiener [8], DS [8], [9] and Enhanced DS (EDS) [8].

## 3 Noise Sources and its Statistical Measurement in Microarray Imaging

It is well known microarray technology can monitor thousand of DNA sequences in a high density array on a glass. The basic procedure for a microarray experiment is simply described as follow. Two mRNA samples are reverse-transcribed into cDNA, labeled using different fluorescent dyes (e.g., the red fluorescent dye Cy5 and the green fluorescent dye Cy3), then mixed and hybridized with the arrayed DNA sequences. After this competitive hybridization, the slides are imaged using a scanner which makes fluorescence measurement for each dye. From the differential hybridization of the two samples, the relative abundance of the spotted DNA sequences can be assessed. A schematic diagram for this process created is shown in [2].

The results of the microarray experiment are two 16-bit tagged mage files, one for each fluorescent dye. The Fig. 3(a) show an example of the mentioned microarray image. As shown in Fig. 3(b), the image is not perfect and it includes noisy sources that blur such image for further gene expression experimentation. The noise source originates from different sources during the course of experiment, such as photon noise, electronic noise, laser light reflection, dust on the slide, and so on. Hence, it is crucial to denoise the resultant image within this process.

Exciting methods to reduce the noise source include using clean glass slide and using a higher laser power rather than higher PMT voltages. However, there are not adequate for the required image qualities and an enhanced software procedure embedded within the process in a much better alternative. In this paper, we focus on the implementation of the SC method (in wavelet domain) to the denoising on microarray images [2]. Yet there are some fundamental obstacles that need clarification before the full potential of microarrays can be explored. One of the major problems in interpretation of microarray data is that different clustering techniques produce different results.

On the other hand, the assessment parameters that are used to evaluate the performance of noise reduction [10], [11] are the following ones:

*Average Absolute Difference (AAD):*

$$AAD = \frac{\sum_{r,c}|I(r,c) - I_d(r,c)|}{R*C} \quad (2)$$

*Signal to Noise Ratio (SNR):*

$$SNR = \frac{\sum_{r,c} I(r,c)^2}{\sum_{r,c}(I(r,c) - I_d(r,c))^2} \quad (3)$$

*Peak Signal to Noise Ratio (PSNR):*

$$PSNR = \frac{R*C*\max_{r,c}(I(r,c)^2)}{\sum_{r,c}(I(r,c) - I_d(r,c))^2} \quad (4)$$

*Image Fidelity (IFy):*

$$\text{IFy} = 1 - \frac{1}{\text{SNR}} \quad (5)$$

*Correlation Quality (CQy):*

$$\text{CQy} = \frac{\sum_{r,c} I(r,c) * I_d(r,c)}{\sum_{r,c} I(r,c)} \quad (6)$$

*Structural Content (SCt):*

$$\text{SCt} = \frac{\sum_{r,c} I(r,c)^2}{\sum_{r,c} I_d(r,c)^2} \quad (7)$$

Where for an image of **R*C** (rows-by-columns) pixels, **r** means row, **c** means column, **I** means original image (without noise), and **$I_d$** means denoised image. Such as, a lower **AAD** gives a "cleaner" image as more noise is reduced; larger **SNR** and **PSNR** indicates a smaller difference between the original (without noise) and denoised image; if **IFy** and **SCt** spread at 1, we will obtain an image **$I_d$** of better quality; and a larger value of **CQy** usually corresponds to a better quantitative performance [10], [11].

On the other hand, to compare edge preservation performances of different noise reduction schemes, we adopt the Pratt's figure of merit (FOM) [9] defined by

$$\text{FOM} = \frac{1}{\max\{\hat{N}, N_{ideal}\}} \sum_{i=1}^{\hat{N}} \frac{1}{1 + d_i^2 \alpha} \quad (8)$$

Where $\hat{N}$ and $N_{ideal}$ are the number of detected and ideal edge pixels, respectively, $d_i$ is the Euclidean distance between the *i*th detected edge pixel and the nearest ideal edge pixel, and $\alpha$ is a constant typically set to 1/9. **FOM** ranges between 0 and 1, with unity for ideal edge detection.

## 4 Results

The simulations demonstrate that the SC technique improves the noise reduction performance to the maximum, for bioimages. Here, we present a set of experimental results using two bioimages. Such images were converted to bitmap file format for their treatment [11]. On the other hand, the statistical filters used inside SC method were DS and EDS. For statistical filters employed along, i.e., Median, Lee, Kuan, Gamma-Map, Enhanced Lee, Frost, Enhanced Frost, Wiener, DS, and EDS, we use a reduction scheme [8].

Figure 3 shows the noisy (30 %) and filtered microarray images used in the first experiment of [1], with a 274-by-274 (pixels) by 65536 (gray levels) bitmap matrix.

Table 1 summarizes the assessment parameters vs. 19 filters for Fig. 3, where En-Lee means Enhanced Lee Filter, En-Frost means Enhanced Frost Filter, ST means Soft-Thresholding, HT means Hard-Thre-sholding and SST means Semi-Soft-Thresholding. The assessment parameters were applied to the whole image.

Figure 4 shows the noisy (10 %) and filtered microarray images used in the second experiment of [1], with a 256-by-256 (pixels) by 65536 (gray levels) bitmap matrix.

Table 2 summarizes the assessment parameters vs. 19 filters for Figure 4.

In both cases, the bioimages were processed by using 10 statistical filters, VisuShrink with Daubechies 4 wavelet basis and 1 level of decomposition (improvements were not noticed with other basis of wavelets) [2], [3], [5], [6], [8], SureShrink, Oracle-Shrink, BayesShrink, NormalShrink, TNN [5]-[8], and SC respectively.

Figures 3 and 4 summarize the edge preservation performance of the SC technique vs. the rest of the filters with a considerably acceptable computational complexity. A 3-by-3 kernel was employed for all statistic noise filters. For TNN [7] the empirical function parameter value $\lambda = 0.01$.

For Lee, Enhanced Lee, Kuan, Gamma, Frost and Enhanced Frost filters the damping factor is set to 1, see [3], [8]. The quantitative results of Table 1 and 2 shows that the SC technique can eliminate noise without distorting useful image information and without destroying the important image edges. Also, in the experiment, the SC outperformed the conventional and no conventional noise reducing filters in terms of edge preservation measured by Pratt figure of merit [9]. In nearly every case in every homogeneous region, the SC produced the lowest standard deviation and was able to preserve the mean value of the region.

The numerical results are further supported by qualitative examination, as shown in Fig. 3 and 4.

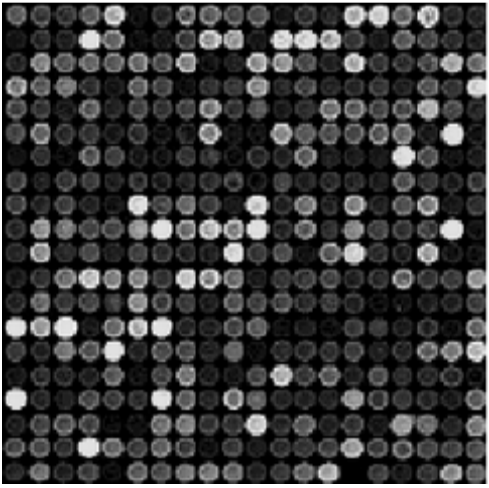
original

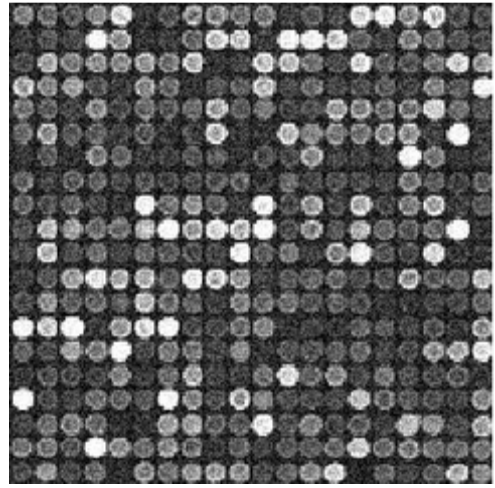
noisy

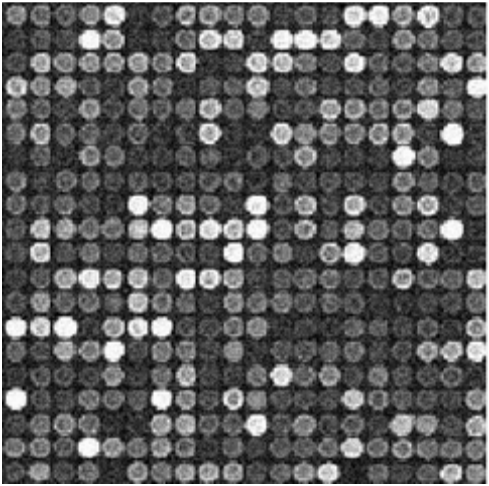
SC

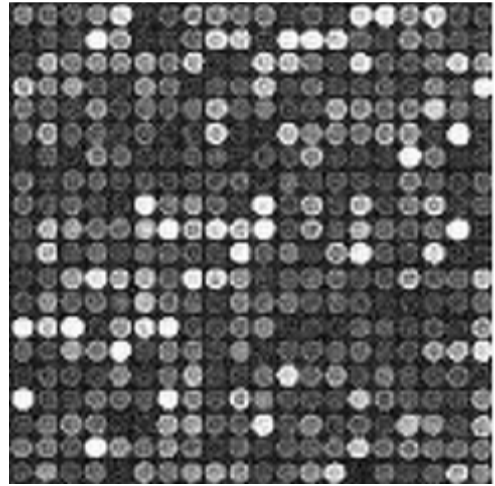
SUREShrink

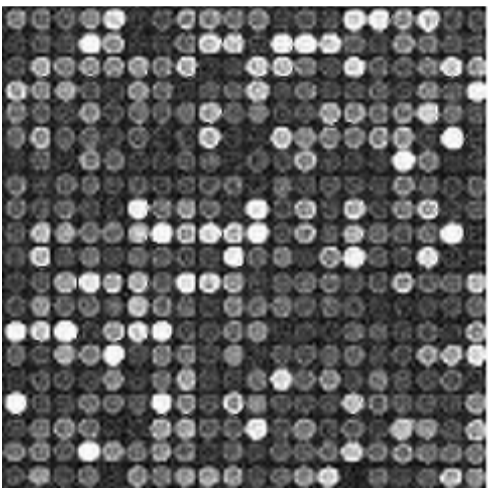
BayesShrink

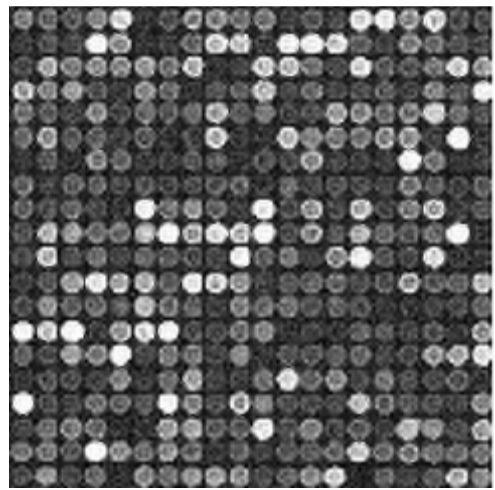
NormalShrink

Fig. 3: Original, noisy and filtered images.

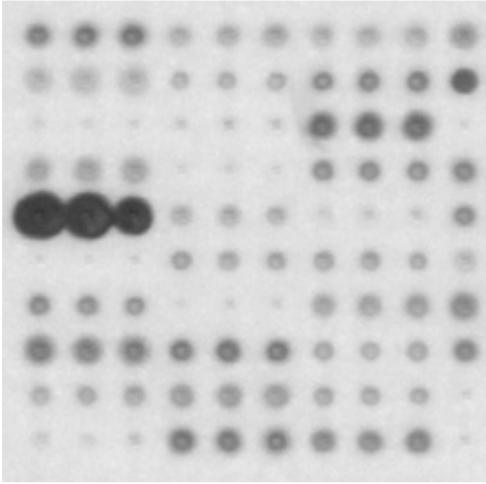
original

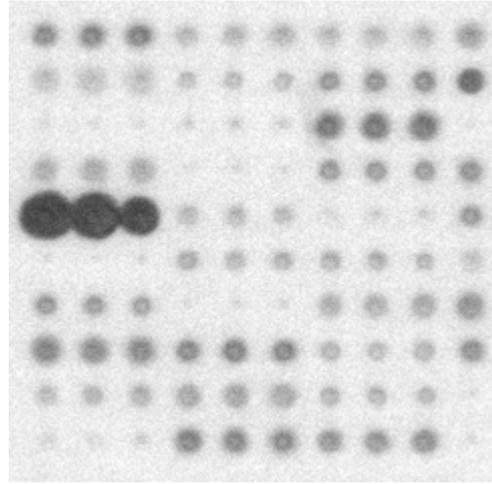
noisy

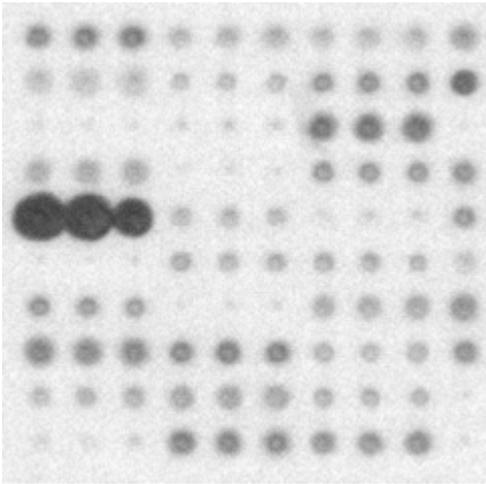
SC

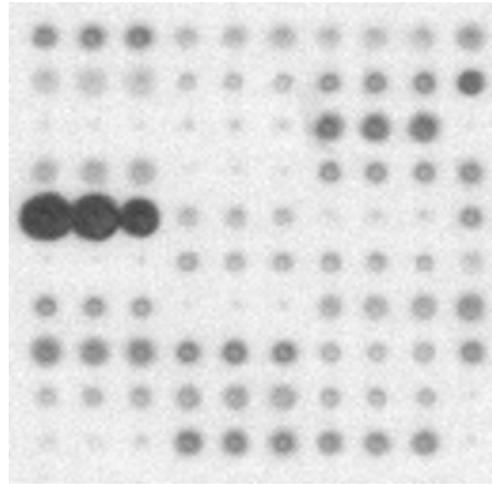
SUREShrink

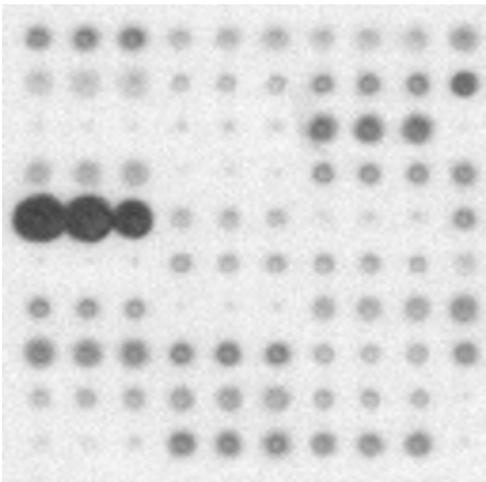
BayesShrink

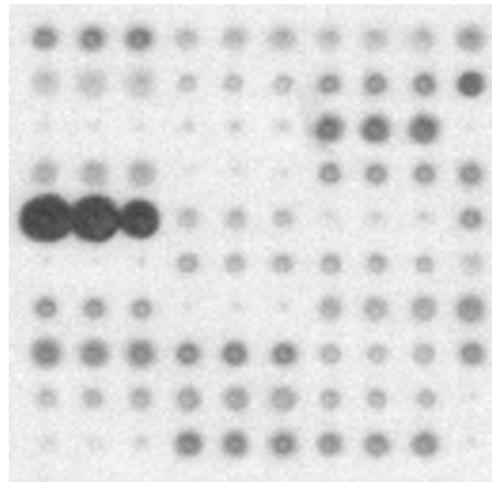
NormalShrink

Fig. 4: Original, noisy and filtered images.

Table 1. Assessment Parameters vs. Filters for Figure 3

| Filter | Assessment Parameter | | | | | | |
|---|---|---|---|---|---|---|---|
| | AAD | SNR | PSNR | IF | CQ | SC | FOM |
| En-Frost | 38.2653 | 3.4464 | 33.7364 | 0.7109 | 150.7467 | 0.5663 | 0.39857 |
| En-Lee | 39.7437 | 3.3363 | 33.8373 | 0.7112 | 150.7472 | 0.5632 | 0.49876 |
| Frost | 38.4374 | 3.2423 | 33.7033 | 0.7106 | 150.5244 | 0.5689 | 0.48756 |
| Lee | 39.2427 | 3.4242 | 32.6363 | 0.7015 | 150.4141 | 0.5924 | 0.43447 |
| Gamma | 39.6252 | 3.1112 | 33.2703 | 0.7063 | 150.1918 | 0.5751 | 0.44235 |
| Kuan | 39.8224 | 3.1243 | 31.8272 | 0.7041 | 149.3121 | 0.5715 | 0.45342 |
| Median | 39.5252 | 3.1131 | 32.7916 | 0.6852 | 148.9172 | 0.5896 | 0.40704 |
| Wiener | 39.1829 | 3.4557 | 33.7033 | 0.7106 | 150.5244 | 0.5689 | 0.44236 |
| DS | 38.7332 | 3.4657 | 33.9997 | 0.7169 | 150.9898 | 0.5599 | 0.64111 |
| EDS | 38.1484 | 3.6969 | 34.1315 | 0.7182 | 151.5252 | 0.5612 | 0.64324 |
| VisuShrink (ST) | 39.1450 | 3.4596 | 33.7412 | 0.7109 | 151.1527 | 0.5657 | 0.44382 |
| VisuShrink (HT) | 38.8612 | 3.5283 | 34.4115 | 0.7166 | 151.3316 | 0.5666 | 0.44324 |
| VisuShrink (SST) | 38.1829 | 3.5557 | 34.7033 | 0.7196 | 151.9202 | 0.5612 | 0.46432 |
| SureShrink | 38.1612 | 3.5751 | 34.7193 | 0.7198 | 151.9244 | 0.5611 | 0.43322 |
| OracleShrink | 38.1189 | 3.6957 | 34.7233 | 0.7198 | 151.9844 | 0.5619 | 0.45534 |
| BayesShrink | 38.1145 | 3.6968 | 34.7237 | 0.7199 | 151.9953 | 0.5612 | 0.46329 |
| NormalShrink | 38.1098 | 3.6998 | 34.8734 | 0.7199 | 151.9983 | 0.5609 | 0.59333 |
| TNN | 38.1008 | 3.7157 | 34.8833 | 0.7199 | 151.9992 | 0.5600 | 0.65432 |
| SC | 37.7155 | 3.7772 | 36.8388 | 0.7353 | 155.4613 | 0.5513 | 0.69123 |

Table 2. Assessment Parameters vs. Filters for Figure 4

| Filter | Assessment Parameter | | | | | | |
|---|---|---|---|---|---|---|---|
| | AAD | SNR | PSNR | IF | CQ | SC | FOM |
| En-Frost | 12.4747 | 290.1324 | 363.6712 | 0.9830 | 226.4744 | 0.8972 | 0.41265 |
| En-Lee | 12.8474 | 290.2522 | 363.9321 | 0.9883 | 226.8373 | 0.8932 | 0.51986 |
| Frost | 12.1847 | 290.2772 | 363.0233 | 0.9828 | 226.3272 | 0.8923 | 0.55312 |
| Lee | 12.3733 | 290.2333 | 363.0238 | 0.9838 | 226.2822 | 0.8943 | 0.44421 |
| Gamma | 12.3830 | 290.8331 | 363.3433 | 0.9882 | 226.8383 | 0.8934 | 0.51235 |
| Kuan | 12.3833 | 290.8272 | 363.4923 | 0.9887 | 226.8381 | 0.8934 | 0.54129 |
| Median | 12.9973 | 289.1212 | 361.8374 | 0.9673 | 225.9287 | 0.8734 | 0.51286 |
| Wiener | 11.9042 | 290.8635 | 363.5568 | 0.9866 | 226.8901 | 0.8954 | 0.56413 |
| DS | 11.4572 | 290.9950 | 363.9393 | 0.9898 | 226.9723 | 0.8993 | 0.64213 |
| EDS | 11.5792 | 290.9998 | 363.9865 | 0.9899 | 226.9975 | 0.8993 | 0.64449 |
| VisuShrink (ST) | 11.9055 | 289.2367 | 361.5523 | 0.9761 | 222.7564 | 0.8872 | 0.51228 |
| VisuShrink (HT) | 11.9042 | 290.8673 | 363.5615 | 0.9966 | 226.8909 | 0.8976 | 0.56424 |
| VisuShrink (SST) | 11.7864 | 290.9546 | 363.9822 | 0.9975 | 226.8937 | 0.8984 | 0.56389 |
| SureShrink | 11.7074 | 291.0753 | 363.8343 | 0.9991 | 226.8942 | 0.8991 | 0.57432 |
| OracleShrink | 11.8436 | 290.9332 | 363.7363 | 0.9968 | 226.8963 | 0.8983 | 0.55234 |
| BayesShrink | 11.9353 | 290.9363 | 363.7361 | 0.9923 | 226.8942 | 0.8962 | 0.56328 |
| NormalShrink | 11.6875 | 290.9992 | 363.9353 | 0.9992 | 226.9021 | 0.8999 | 0.59611 |
| TNN | 11.4447 | 291.7243 | 363.9991 | 0.9994 | 226.9732 | 0.9002 | 0.62900 |
| SC | 10.9071 | 294.9237 | 383.1090 | 0.9992 | 229.8972 | 0.9173 | 0.69322 |

On the other hand, all filters was applied to complete image, for Figure 3 (274-by-274) pixels and Figure 4 (256-by-256) pixels, and all the filters were implemented in MATLAB® (Mathworks, Natick, MA) on a PC with an Athlon (2.4 GHz) processor.

## 5 Conclusion

In this paper we have developed a SC technique based tools for removing additive noise in microarrays. The simulations show that the SC have better performance than the most commonly used filters for microarrays (for the studied benchmark parameters) which include statistical filters, wavelets, and a version of TNN. The SC exploits the local coefficient of variations in reducing noise. The performance figures obtained by means of computer simulations reveal that the SC technique provides superior performance in comparison to the above mentioned filters in terms of smoothing uniform regions and preserving edges and features. The effectiveness of the technique encourages the possibility of using the approach in a number of ultrasound and radar applications. Besides, the method is computationally efficient and can significantly reduce the noise while preserving the resolution of the original microarray image. Considerably increased Pratt's figure of merit strongly indicates improvement in detection performance. Also, cleaner images suggest potential improvements for classification and recognition.

*References:*
[1] E.C. Rouchka. (2004, April). Lecture 12: Microarray Image Analysis. [Online]. Available: http://kbrin.a-bldg.louisville.edu/CECS694/Lecture12.ppt
[2] X.H. Wang, S.H. Istepanian, and Y.H. Song, "Microarray Image Enhancement by Denoising Using Stationary Wavelet Transform," *IEEE Transactions on Nanobioscience*, vol.2, no. 4, pp.184-189, December 2003. [Online]. Available: http://technology.kingston.ac.uk/momed/papers/IEEE_TN_Micorarray_Wavelet%20Denoising.pdf
[3] H.S. Tan. (2001, October). Denoising of Noise Speckle in Radar Image. [Online]. Available: http://innovexpo.itee.uq.edu.au/2001/projects/s804294/thesis.pdf
[4] H. Guo, J.E. Odegard, M. Lang, R.A. Gopinath, I. Selesnick, and C.S. Burrus, "Speckle reduction via wavelet shrinkage with application to SAR based ATD/R," Technical Report CML TR94-02, CML, Rice University, Houston, 1994.
[5] D.L. Donoho and I.M. Johnstone, "Adapting to unknown smoothness via wavelet shrinkage," *Journal of the American Statistical Association*, vol. 90, no. 432, pp. 1200-1224, 1995.
[6] S.G. Chang, B. Yu, and M. Vetterli, "Adaptive wavelet thresholding for image denoising and compression," *IEEE Transactions on Image Processing*, vol. 9, no. 9, pp.1532-1546, September 2000.
[7] X.-P. Zhang, "Thresholding Neural Network for Adaptive Noise reduction," *IEEE Trans. on Neural Networks*, vol.12, no. 3, pp.567-584, May 2001.
[8] M. Mastriani and A. Giraldez, "Enhanced Directional Smoothing Algorithm for Edge-Preserving Smoothing of Synthetic-Aperture Radar Images," *Journal of Measurement Science Review*, vol 4, no. 3, pp.1-11, 2004. [Online]. Available: http://www.measurement.sk/2004/S3/Mastriani.pdf
[9] Y. Yu, and S.T. Acton, "Speckle Reducing Anisotropic Diffusion," *IEEE Trans. on Image Processing*, vol. 11, no. 11, pp.1260-1270, 2002.
[10] G. Delfino and F. Martinez. (2000, March). Watermarking insertion in digital images (spanish). [Online]. Available: http://www.internet.com.uy/fabianm/watermarking.pdf
[11] A.K. Jain, *Fundamentals of Digital Image Processing,* Englewood Cliffs, New Jersey, 1989.

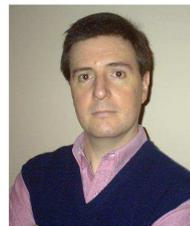
**Prof. Mario Mastriani** was born in Buenos Aires, Argentina on February 1, 1962. He is Professor of Digital Signal Processing and Digital Image Processing of National Technological University (UTN). He is working in the speckle filter of the first synthetic aperture radar (SAR) satellite of SAOCOM Mission, National Commission of Space Activities (CONAE), Buenos Aires, Argentina. He published 33 papers. He is a currently reviewer of IEEE Transactions on Neural Networks, Signal Processing Letters, Transactions on Image Processing, Transactions on Signal Processing, Communications Letters, Transactions on Geoscience and Remote Sensing, Transactions on Medical Imaging, Transactions on Biomedical Engineering, Transactions on Fuzzy Systems; and Springer-Verlag Journal of Digital Imaging.
He (M'05) became a member (M) of ENFORMATIKA in 2004. His areas of interest include Digital Signal Processing, Digital Image Processing, wavelets and Neural Networks.

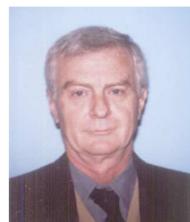
**Prof. Alberto E. Giraldez** was born in Trelew, Chubut, Argentina on January 2, 1949. He received his B.S. degree, and Ph.D. in Physics in 1971 and 1975 respectively, from La Plata, University, Argentina. He is Professor of Electromagnetism of Buenos Aires Institute of Technology (ITBA). He is the scientific manager of the first synthetic aperture radar (SAR) satellite of SAOCOM Mission, National Commission of Space Activities (CONAE), Buenos Aires, Argentina. He published 70 papers. He was a reviewer of Radio Science (American Geophysical Union), and Annales Geophysicae (European Geophysical Society). He is the manager of the Ionosferic Laboratory (LIARA) of Argentine Navy from 1980. He is the president of the work interim group 6/8 of the International Radiocommunication Consultive Committee (IRCC) of the International Telecommunication Union (ITU) elect in 1978, and with command renovated in 1982 and 1986 for the study of "Anomalous propagation in Very High Frequency for reflections in the Ionosphere". His areas of interest include Synthetic Aperture Radar, Digital Signal Processing, Radiopropagation and Electromagnetic Compatibility, and Radar Technology.